\documentclass[aps,twocolumn,prl,tightenlines,floatfix,showpacs]{revtex4}
\usepackage{graphicx}
\usepackage[english]{babel}
\usepackage{amsmath}
\usepackage{amssymb}
\usepackage{times}

\begin{document}
\title{Dynamical crossover between the infinite-volume and empty-lattice limits of ultra-cold fermions in 1D optical lattices}
\author{Chih-Chun Chien$^{1}$ and Massimiliano Di Ventra$^{2}$}

\affiliation{$^{1}$Theoretical Division, Los Alamos National Laboratory, MS B213, Los Alamos, NM 87545, USA \\
$^{2}$Department of Physics, University of California, San Diego, CA 92093, USA
}
\date{\today}

\begin{abstract}
Unlike typical condensed-matter systems, ultra-cold atoms loaded into optical lattices allow separate control of {\it both} the particle number {\it and} system size.
As a consequence, there are two distinct "thermodynamic" limits that
can be defined for these systems: {\it i)} "infinite-volume limit" at constant finite density, and {\it ii)} "empty-lattice limit" at constant particle number. To probe the difference between these two limits and their crossover, we consider a partially occupied lattice and study the transport of non-interacting fermions and fermions interacting at the mean-field level into the unoccupied region. In the infinite-volume limit, a finite steady-state current emerges. On the other hand, in the empty-lattice limit there is no finite steady-state current. By changing the initial filling, we find a smooth crossover between the two limits. Our predictions may be verified using available experimental tools and demonstrate a fundamental difference between isolated small systems such as ultra-cold atoms and conventional
condensed-matter systems.
\end{abstract}


\maketitle

In statistical physics, the thermodynamic limit refers to fixing the density $N_p/V$ - where $N_p$ is the particle number and $V$ is the volume - and let $V\rightarrow\infty$ for a continuum system, or fixing $N_p/N$
- where $N$ is the number of lattice sites - and taking $N\rightarrow\infty$ for a lattice system \cite{Todabook}. The thermodynamic limit is important in the correct description of phase transitions, extensiveness of statistical quantities, Bose-Einstein condensation, and many other physical phenomena (see, e.g., a recent review in Ref.~\cite{TDlimit_review}).

In typical condensed-matter systems one cannot arbitrarily increase the volume (or number of lattice sites) without simultaneously increasing the particle number. As a consequence, one is led to define a {\it unique} thermodynamic limit. This is not the case for ultra-cold atomic systems, where  both $N_p$ and $V$ (or $N$) \textit{are separately controllable}. In this case, we can then define two {\it distinct} thermodynamic limits: one - that we may call "infinite-volume limit" (IVL) - where we fix the density while increasing the system size, and the other - we call "empty-lattice limit" (ELL) - where we keep the particle number constant while increasing the system size. The ELL has been less explored in conventional solid-state systems, but it is of fundamental importance in statistical physics and may have applications in ultra-cold atomic systems \cite{MIT_polaron,Jochim11}. In the ELL the Fermi energy is close to the bottom of the energy band due to the small number of particles compared to the large number of available energy states. Therefore, the physical properties in these two limits may be quite different.

A concrete example of this difference in ultra-cold atoms is the strongly-interacting spin-imbalanced (polarized) Fermi gas \cite{ZSSK06,Rice1,MIT_polaron}. Its stable structure at low temperatures has been proposed, based on many-body theories in the IVL, to be either a polarized superfluid or a separated structure of a superfluid and a normal phase \cite{ChienPRL}. This agrees with experimental results with finite ratios between the population of the two hyper-fine states \cite{ZSSK06,Rice1}. However, when \textit{one single} fermion is immersed in a cloud of fermions of the opposite spin, it behaves like a polaron and exhibits very different physics \cite{MIT_polaron}. Due to difficulties in formulating the problem with strong interactions, the crossover between the single-impurity polaron and the finite-population-ratio superfluid remains an open question. Finding another case where the transition
between the two limits can be easily checked, both experimentally and theoretically, would be thus desirable.

In this respect, ultra-cold atomic systems present several aspects that differ significantly from those of conventional condensed-matter systems, which make them ideal candidates to explore such fundamental issues. For one, the interactions in ultra-cold atoms can be {\it turned off}, thus allowing one to experimentally study the role of interactions in physical phenomena in a controlled way \cite{BlochRMP}. This is of particular importance in investigating non-equilibrium properties because, due to Coulomb interactions, finding a genuine non-interacting system is difficult in conventional electronic systems. In addition, it is clearly discussed in Ref.~\cite{Ziff_IBG} that for a non-interacting Bose gas with Bose-Einstein condensation (BEC), the prediction of the number fluctuations of the condensate as obtained from the canonical ensemble does not agree with that from the grand-canonical ensemble. This difference is based on the fundamental assumption of whether there are exchanges of particles with a reservoir, and may be tested in recent experiments of trapped atomic clouds, which to all practical purposes can be viewed as isolated systems \cite{Chin_fluctuations}.

Here we explore the crossover between the IVL and ELL by focusing on one observable: the formation of quasi steady-state currents in one-dimensional (1D) atomic gases set out of equilibrium in optical lattices. We will focus on non-interacting fermions and fermions interacting at the mean-field level. It was recently shown that when these ultra-cold fermions are loaded into optical lattices and set out of equilibrium, a quasi steady-state current (QSSC) - characterized by a current plateau as a function of time - may be observed \cite{MCFshort}. In the IVL this QSSC then develops into a true global steady state. We can however ask how such a fermionic current evolves from the IVL to the ELL and vice versa.

For a single fermion in a lattice of size $N$, we will present a general argument which rules out a finite average current as $N\rightarrow\infty$. For the setup discussed in this paper, and for non-interacting fermions, we show explicitly that the current decreases as $N^{-3/2}$. On the other hand, for $N_p/N=O(1)$ on a lattice of reasonable size, a QSSC is readily observable in several different setups \cite{Bushong05,MCFshort}. By changing $N_p$ and $N$, the QSSC vanishes in a continuous fashion, and no sharp transition is found. A finite steady-state current is therefore a many-particle phenomenon, not a property of few particles, at least for the initial conditions 
and setup considered in this work.

We begin with a one-dimensional finite lattice of size $N$ and consider single-species fermions first. Due to Pauli exclusion principle, the $s$-wave interactions between fermions are suppressed so there is virtually no interaction in the system. We model the fermions in the lattice by a
 tight-binding Hamiltonian
\begin{equation}\label{eq:H}
H=-\tilde{t}\sum_{\langle ij\rangle}c^{\dagger}_{i}c_{j}.
\end{equation}
Here $\langle ij\rangle$ denotes nearest-neighbor pairs, $\tilde{t}$ is the tunneling coefficient, and $c_i^{\dagger}$ ($c_i$) is the creation (annihilation) operator of site $i$. The unit of time is $t_0\equiv\hbar/\tilde{t}$.  Experimentally, such a 1D lattice may be realized by inserting an optical barrier \cite{ring1} into a ring of optical lattices \cite{ring2}. The resulting C-shaped lattice is geometrically identical to an open 1D lattice. Another possible technique is the spatial light modulation \cite{SLM2003,SLM2008} which can produce designed patterns of trap potentials.

\begin{figure}
  \includegraphics[width=2in,clip]
{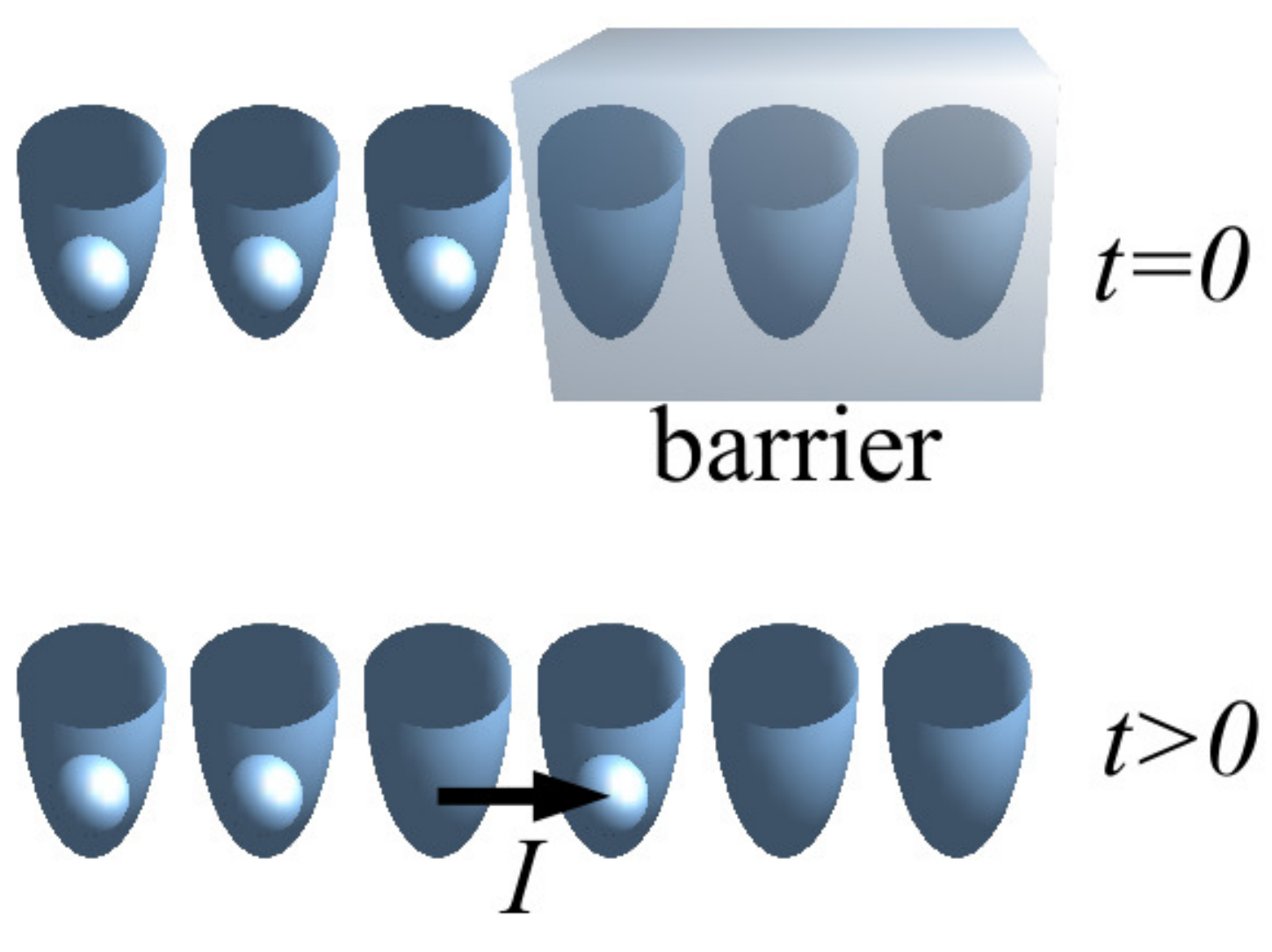}
  \caption{Schematic plot of the setup. The dashed box denotes a barrier at $t=0$ so particles can only populate the left half lattice. The barrier is then lifted and a current ensues. Here particles may be in a superposition of quantum states.}
\label{fig:TD_cartoon}
\end{figure}

Figure~\ref{fig:TD_cartoon} illustrates our proposed experimental setup. Initially there is a barrier blocking particles from entering the right half of the lattice and the atoms are in the ground state of the left half lattice. The system is then driven out of equilibrium by removing the barrier. As the atoms move to the right, a current through the middle of the lattice develops. Since the system we consider is finite and closed, we employ the microcanonical formalism (MCF)~\cite{Micro,Maxbook} as implemented in Ref.~\cite{MCFshort}.

In the MCF one monitors the evolution of the correlation matrix $C(t)$, whose elements are $c_{ij}(t)=\langle S_0|c^{\dagger}_{i}(t)c_{j}(t)|S_0\rangle$. Here $|S_0\rangle$ denotes the initial quantum state. By using a unitary transformation $c_{j}=\sum_{k}(U)_{jk}d_{k}$, one can rewrite $H$ as $H=\sum_{p}\epsilon^{e}_{p}d^{\dagger}_{p}d_{p}$. Here $\epsilon^{e}_{p}$ denotes the energy spectrum of $H$. Explicitly for the Hamiltonian shown in Eq.~\eqref{eq:H}, $U_{jp}=\sqrt{\frac{2}{N+1}}\sin(\frac{jp\pi}{N+1})$ and $\epsilon^{e}_{p}=-2\tilde{t}\cos(\frac{p\pi}{N+1})$. The index $p$ denotes the energy level and should not be confused with the momentum. From the equation of motion $i(dc_{j}(t)/dt)=[c_j(t),H]$ it follows that $c_j(t)=\sum_{p}(U)_{jp}d_{p}(0)\exp(-i\epsilon^{e}_{p}t)$ ($\hbar=1$ throughout the paper). Then the time evolution is given by
\begin{eqnarray}\label{eq:cij}
c_{ij}(t)&=&\sum_{p,p^\prime=1}^{N}(U^{\dagger})_{pi}(U)_{jp^\prime}D_{pp^\prime}(0)e^{i(\epsilon^{e}_{p}-\epsilon^{e}_{p^{\prime}})t}.
\end{eqnarray}
Here $D_{pp^\prime}(0)=\sum_{i,j}(U^{\dagger})_{ip}(U)_{p^\prime j}c_{ij}(t=0)$ is the initial correlation matrix in the energy basis.

The current flowing from the left to the right for one species is $I=-\langle d\hat{N}_L(t)/dt\rangle$, where $\hat{N}_L(t)=\sum_{i=1}^{N/2}c^{\dagger}_{i}(t)c_{i}(t)$. For the Hamiltonian considered here,
\begin{equation}\label{eq:I}
I(t)=2\tilde{t}\mbox{Im}\{c_{N/2,N/2+1}(t)\}.
\end{equation}
Experimentally one may prepare several identical setups and take density images at different times. The current corresponds to the rate at which atoms are tranferred to the right.

We consider the transport of $N_p$ single-species fermions in a lattice of size $N$ and the filling factor is defined as $n=N_p/N$. The initial state has the lowest $N_p$ energy states of the left half Hamiltonian $H_{L}=-\tilde{t}\sum_{1\le \langle ij\rangle \le N/2}c^{\dagger}_{i}c_{j}$ occupied, while the right half lattice is empty. In the energy basis of $H_{L}$, the correlation function is $D_{pp^{\prime}}^{L}=\theta(N_p-p)\delta_{pp^{\prime}}$, where $\theta(x)=1$ if $x\ge 0$ and $\theta(x)=0$ otherwise. This corresponds to a Fermi sea of $N_p$ particles on the left half lattice with Fermi energy $E_F=-2\tilde{t}\cos[\frac{N_p \pi}{(N/2)+1}]$. The initial correlation function in real space is thus $c_{ij}(t=0)=\sum_{p,p^\prime =1}^{N_p}(U_{L}^{\dagger})_{pi}(U_{L})_{jp^\prime}D^{L}_{pp^\prime}$ if $1\le i,j \le (N/2)$ and zero otherwise, where $(U_{L})_{jp}=\sqrt{\frac{4}{N+2}}\sin(\frac{2jp\pi}{N+2})$ is the unitary transformation for $H_L$. The time-evolved correlation matrix is given by Eq.~\eqref{eq:cij} and the corresponding current follows Eq.~\eqref{eq:I}.

\begin{figure}
  \includegraphics[width=3.4in,clip]
{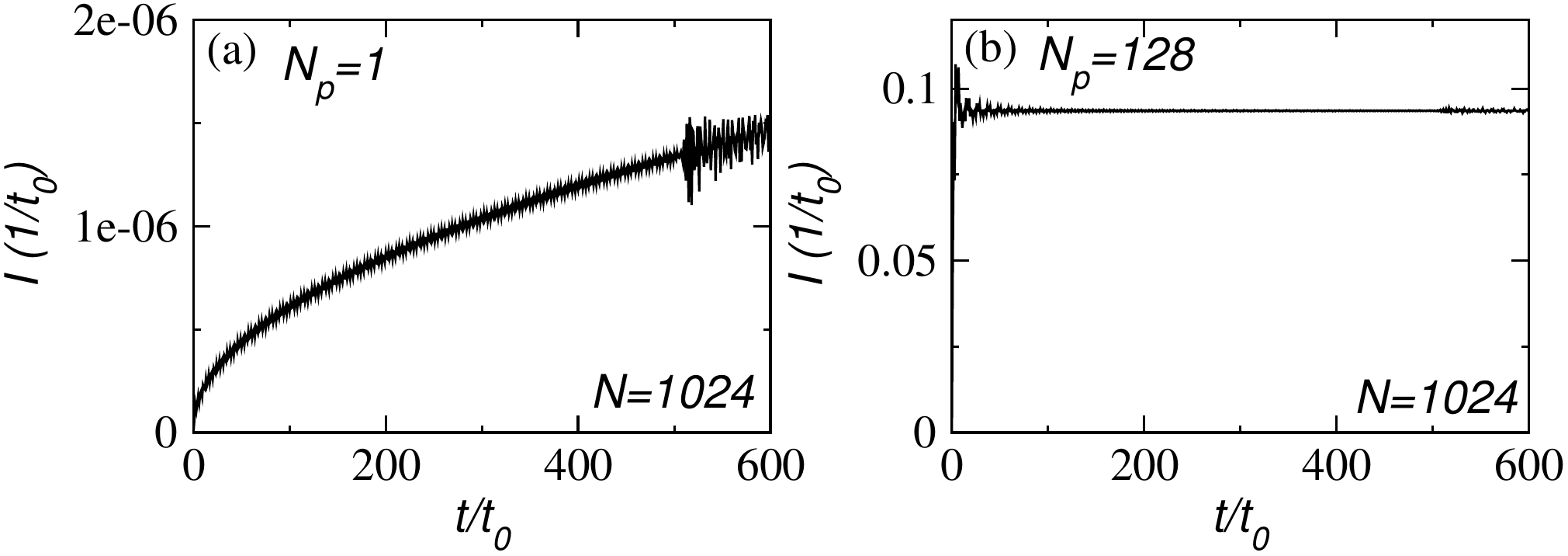}
  \caption{The currents of non-interacting fermions for (a) $N_p=1$ and (b) $N_p=128$ in a lattice of size $N=1024$.}
\label{fig:I_contrast}
\end{figure}
Figure~\ref{fig:I_contrast} shows the currents from one particle ($N_p=1$) and $128$ particles in a lattice of size $N=1024$. Due to finite size effects, there is revival behavior when $t>(N/2)t_0$ and we will focus on the physics before the first revival occurs. In other words, we focus on the finite-time behavior ($t\le (N/2)t_0$) \textit{before} taking the $N\rightarrow\infty$ limit. In the opposite order of limits, $t\rightarrow\infty$ before $N\rightarrow\infty$, the system goes through many revivals, and the discussion of the QSSC is not meaningful.

There is a stark difference in the currents of the two cases. A QSSC corresponding to a plateau in the current as a function of time can be observed in the $N_p=128$ case. Similar QSSCs have been reported in many different setups \cite{Bushong05,MCFshort} with filling of $O(1)$. In contrast, the current of the single-particle case is extremely small and does not exhibit a QSSC. The origin of this difference is closely related to how the $N\rightarrow\infty$ limit is approached. As anticipated, when both $N_p$ and $N$ are tunable, one can {\it i)} fix the ratio of $n=N_p/N$ as $N\rightarrow\infty$, which corresponds to the IVL, or {\it ii)} fix $N_p$ as $N\rightarrow\infty$ so that $n\rightarrow 0$, which corresponds to the ELL.

We first prove analytically that for the single-particle case the current eventually decays to zero as $N\rightarrow\infty$. From Eqs~\eqref{eq:cij} and \eqref{eq:I}, for $N_p=1$ we have
\begin{eqnarray}
I&=&-2\tilde{t}\sum_{p\neq p^{\prime}}\sin\left\{2\tilde{t}t\left[\cos(\frac{p\pi}{N+1})-\cos(\frac{p^{\prime}\pi}{N+1})\right] \right\}\times \nonumber \\
& &U_{N/2,p}U_{N/2+1,p^{\prime}}F(p)F(p^{\prime}).
\end{eqnarray}
Here $F(p)=\sum_{j=1}^{N/2}U_{L,j1}U_{jp}$. In the limit $N\rightarrow\infty$, it becomes $F(p)\rightarrow 2\sqrt{2}\int_{0}^{1/2}dx[\sin(2\pi x)\sin(\pi xp)]$ or explicitly, $F(p)\rightarrow 4\sqrt{2}\sin(\frac{p\pi}{2})/[(4-p^{2})\pi]$. Therefore $F(p)=0$ if $p$ is even, except $F(2)=\sqrt{2}/2$, and $F(p)$ is finite for $p$ odd. In the final expression, all terms with $\{p=\textrm{odd}, p^{\prime}=\textrm{odd}\}$ cancel each other so only terms with $\{p=2, p^{\prime}=\textrm{odd}\}$ and $\{p^{\prime}=2,p=\textrm{odd}\}$ contribute and their contributions are equal. Thus we consider the contribution from $\{p=2, p^{\prime}=\textrm{odd}\}$ and the final result is twice of this contribution. To make the analysis explicit, we focus on the long-time limit and set $t=(N/2)t_0$. One can make other choices as long as $t/t_{0}\sim O(N/2)$ and obtain the same conclusion. Due to the $\sin$ function,  the first term oscillates rapidly for large $t/t_0$. For $p=2$, only $0<p^{\prime}/(N+1)<1/\sqrt{N}$ may contribute finitely. The current then approaches
\begin{eqnarray}
I&\rightarrow&\frac{8\tilde{t}}{2\pi}\int_{0}^{\frac{1}{\sqrt{N}}}d\bar{p}\sin[N(1-\cos(\bar{p}\pi))]\frac{\pi}{N+1}\frac{2}{\bar{p}^{2}\pi} \propto N^{-\frac{3}{2}} \nonumber
\end{eqnarray}
Therefore as $N\rightarrow\infty$ (in the ELL), the current $I\rightarrow 0$.

On the other hand, a finite QSSC in the IVL requires $n=O(1)$. This can be seen by observing that for a finite QSSC with an averaged value $I_s$ to exist during a time interval, say from $t=(N/4)t_0$ to $t=(N/2)t_0$, the transmitted particles must be $\Delta N_{L}=\int_{(N/4)t_0}^{(N/2)t_0}I(t)dt\approx (N/4)t_0 I_s =O(N)$. Since $\Delta N_{L}\le N_p$, we conclude that a finite QSSC requires $n=N_p/N=O(1)$ as $N\rightarrow \infty$, and this is the conventional thermodynamic limit in many condensed-matter settings. We remark that in this argument one can take any time interval for the QSSC as long as it is of order $O(N)t_0$, since one expects a QSSC to last for a macroscopic time scale. The linear scaling of the time duration where a QSSC can be observed with respect to the lattice size has been reported in Refs.~\cite{Bushong05,MCFshort}. After presenting the results of the current in the two different limits, we now study its behavior in between those two limits.

\begin{figure}
  \includegraphics[width=3.4in,clip]
{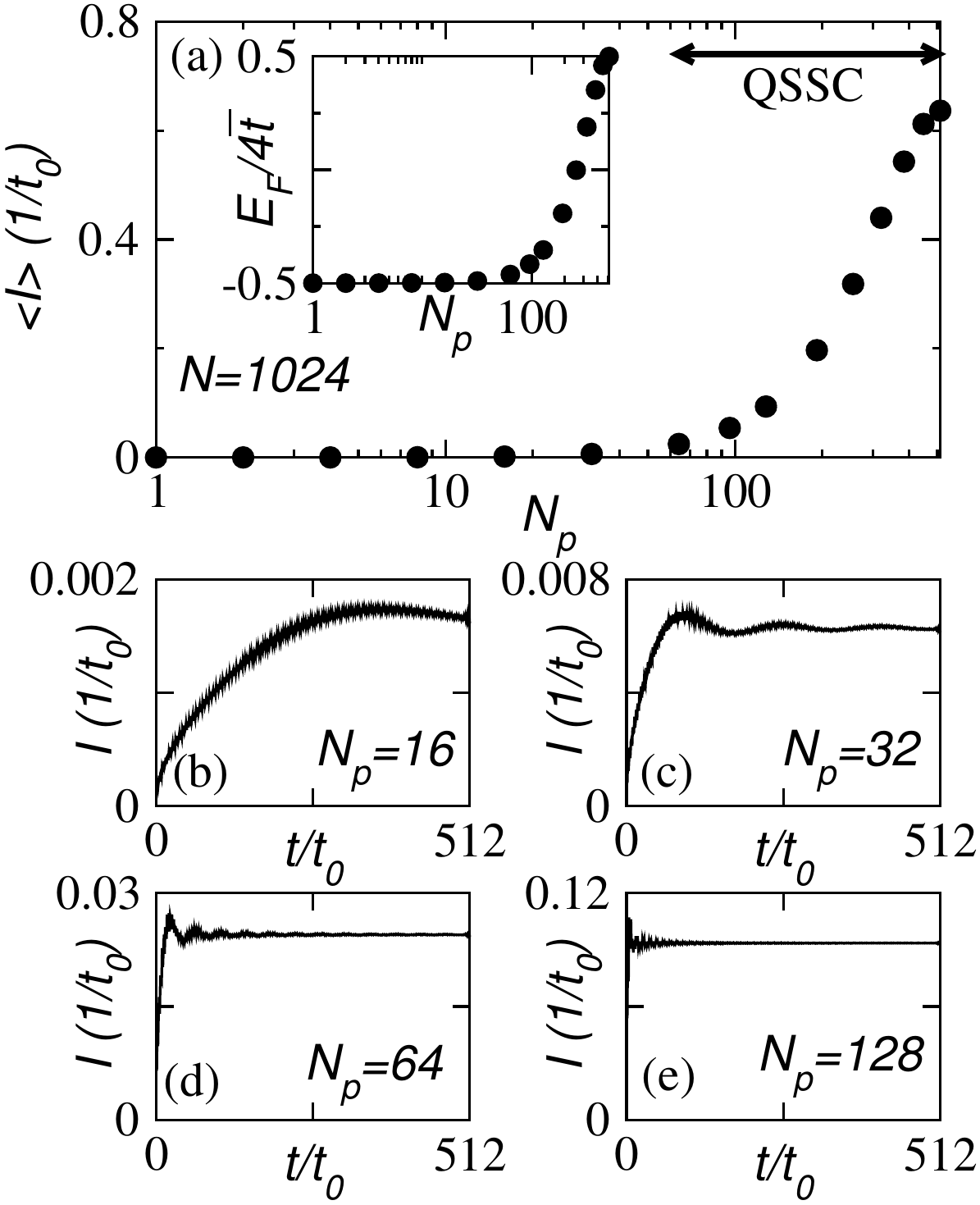}
  \caption{(a) The averaged current (defined in Eq.~\eqref{eq:Iavg}) as a function of particle number $N_p$ in a lattice of size $N=1024$. A QSSC can be found when $N_p\ge 64$. Inset: Normalized Fermi energy $E_F/4\tilde{t}$ as a function of $N_p$ with $N=1024$. The current as a function of time for (b) $N_p=16$, (c) $N_p=32$, (d) $N_p=64$, and (e) $N_p=128$ is shown.}
\label{fig:IvsNp}
\end{figure}

{\it From the ELL to the IVL -} We first fix the lattice size $N$ and study the effect of varying $N_p$. In order not to be distracted by small oscillations in the QSSC, we define the averaged current as
\begin{equation}\label{eq:Iavg}
\langle I\rangle\equiv\frac{1}{(N/4)t_0}\int_{T_0}^{T_0+(N/4)t_0}I(t) dt.
\end{equation}
In our simulations we chose $T_0=[(N/4)-3]t_0$ to avoid the few points around $(N/2)t_0$ where the current starts to exhibit boundary effects. When there is a QSSC - like the one shown in Fig.~\ref{fig:I_contrast}(b) - $\langle I\rangle$ gives the magnitude of the QSSC. When there is no QSSC, $\langle I\rangle$ gives an estimate of the averaged current. Figure~\ref{fig:IvsNp} shows $\langle I\rangle$ as a function of $N_p$ for fixed $N=1024$. Due to our selection of the initial ground state, $1\le N_p\le 512$.

There is no significant current for $N_p\le16$ so this regime is basically in the ELL. A QSSC can be observed for $N_p\ge 64$ and we indicate the regime where a QSSC can be found. Therefore for $N_p\ge 64$ the system enters the IVL  because $n=O(1)$. The vanishing of the QSSC in the transient regime $16\le N_p\le 64$ is a slow crossover. Here, the oscillations in the current become visible and a plateau of length of $O(N/4)t_0$ can no longer be identified. We demonstrate this crossover by showing the currents (not the averaged ones) for $N_p=16, 32, 64, 128$ in Figure~\ref{fig:IvsNp}. The Fermi energy $E_F$ normalized to $4\tilde{t}$ as a function of $N_p$ is also shown in the inset of Fig.~\ref{fig:IvsNp}. As expected, in the ELL ($N_p\le 16$) the Fermi energy is close to the bottom of the energy band while in the IVL ($N_p\ge 64$) the Fermi energy is above the bottom of the energy band. Thus one can see that a finite QSSC emerges when $E_F$ starts to deviate from the bottom of the energy band.

\begin{figure}
  \includegraphics[width=3.4in,clip]
{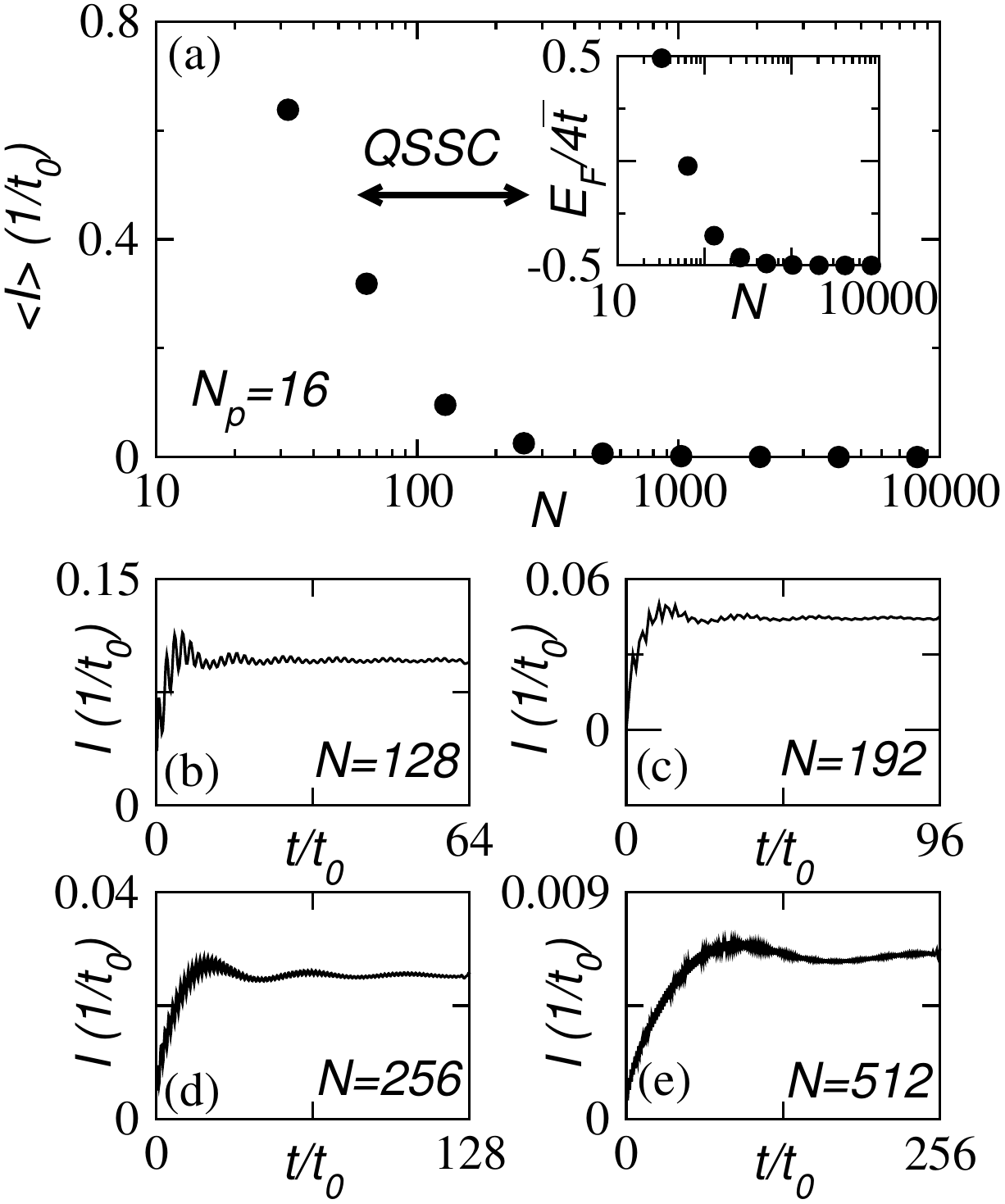}
  \caption{(a) The averaged current (defined in Eq.~\eqref{eq:Iavg}) as a function of $N$ for fixed $N_p=16$. A finite QSSC can be found when $64\le N\le 128$. Inset: Normalized Fermi energy $E_F/4\tilde{t}$ as a function of $N$ with $N_p=16$. The current as a function of time for (b) $N=128$, (c) $N=192$, (d) $N=256$, and (e) $N=512$ is also shown here.}
\label{fig:IvsNSITE}
\end{figure}
{\it From the IVL to the ELL -} We can, however, follow the opposite route, namely fix the particle number $N_p$ and vary the lattice size $N$. Figure~\ref{fig:IvsNSITE} shows the current at fixed $N_p=16$ and different $N$. Here, we report $32\le N\le 8192$. We find that the current becomes negligible for $N\ge 1024$ and the system is in the empty-lattice limit.  Since the IVL practically requires $N\gg 1$, the system in such a setup cannot reach the thermodynamic limit with finite filling. However, a QSSC with a plateau of length $O(N/4)t_0$ can be observed for $64\le N\le 128$. When $N=32$, finite-size effects are strong enough that no plateau in the current can be found. Our findings thus suggest that although a QSSC should be found for $N\gg 1$ in the most strict sense, it already exists for finite systems with reasonable size at filling of order $O(1)$. To clarify how the current of the QSSC vanishes as $N$ increases, the currents for $N=128, 192, 256, 512$ with $N_p=16$ are shown in Figure~\ref{fig:IvsNSITE}. The normalized Fermi energy $E_F/(4\tilde{t})$ as a function of $N$ is shown in the inset of Fig.~\ref{fig:IvsNSITE}. Again, one can see that when $E_F$ deviates from the bottom of the energy band, a finite QSSC starts to form.

After discussing the IVL and ELL separately, we briefly compare the threshold value of the filling factor $n$ for obtaining a finite current in these two different limits. It is important to check that as $n$ increases, a finite QSSC starts to emerge when $n\sim 1/16$ as one can see from Figs.~\ref{fig:IvsNp} and \ref{fig:IvsNSITE}. This is based on the premise that the size $N$ should be large enough ($N>32$ in our simulations). The threshold $1/16$ corresponds to where $E_F$ starts to deviate from the bottom of the energy band, and it indicates that the system is
large enough to transition from the ELL to the IVL. Although the ratio $1/16$ may be specific to the setup discussed here, a threshold with a finite $N_p/N$ should be a generic feature for a finite QSSC to be observable. The finite-size effect introduces another threshold for the ELL case when $N$ becomes too small. Indeed, we observe no plateau in the current in the ELL case, Fig.~\ref{fig:IvsNSITE}, when $N<64$ even though $n>1/4$. The current shows oscillating behavior when $N<64$ due to finite-size effects. To summarize, there is a common threshold - in correspondence to the Fermi energy deviating from the bottom of the energy band - at $n\sim 1/16$ when one tunes either the number of particles or the number of lattice sites. However, for the case of decreasing $N$ from the ELL limit, there is another threshold when $N$ is too small so there is not enough time for a QSSC to develop due to finite-size effects.

Before closing our discussions of the QSSC in non-interacting fermions, we comment on whether it can last long enough for experimental observations. The time scale $t_0$ is estimated to be of the order of milliseconds in present experiments \cite{MCFshort}. Therefore the plateaus shown in Figs.~\ref{fig:IvsNp} and \ref{fig:IvsNSITE} should be observable since they last for at least tens of $t_0$. In Ref.~\cite{MCFshort} it has been shown that for non-interacting fermions the QSSC is also robust against an additional harmonic potential and finite temperature effects, so it may be studied systematically in experiments.


{\it The role of mean-field interactions -}
Next we address the important issue whether the QSSC in the IVL can survive when there are interactions among fermions. Here we consider two-species fermions interacting via weak $s$-wave scattering and model the system as the conventional Hubbard model
\begin{eqnarray}\label{eq:He}
H_{int}&=&H+\sum_{i=1}^{N}U\hat{n}_{i\sigma}\hat{n}_{i\bar{\sigma}}.
\end{eqnarray}
Here $\hat{n}_{i\sigma}=c_{i\sigma}^{\dagger}c_{i\sigma}$, $U$ is the onsite repulsive coupling constant, and $\bar{\sigma}$ is the opposite of $\sigma$. This model should be appropriate for moderate lattice depth \cite{Hubbard_model} and the onsite interaction may be generated by tuning the system near a Feshbach resonance \cite{BlochRMP}. We assume there are $N_{p\sigma}$ fermions of each species on the lattice and define the filling as $N_{p\sigma}/N$.

In the presence of interactions, finding the complete set of eigenstates for $H_{int}$ for a moderate size of lattices becomes difficult. Therefore, instead of writing down the time-evolved correlation matrix in terms of the time-evolved energy eigenstates, we evaluate the correlation matrix by solving the equations of motion. Explicitly, $i(\partial \langle c^{\dagger}_{i\sigma}c_{j\sigma}\rangle/\partial t)=\langle[c^{\dagger}_{i\sigma},H_{e}]c_{j\sigma} \rangle+\langle c^{\dagger}_{i\sigma}[c_{j\sigma},H_{e}]\rangle$, where $[\cdot,\cdot]$ denotes the commutator of the corresponding operators. One obtains
\begin{eqnarray}
i\frac{\partial \langle c^{\dagger}_{i\sigma}c_{j\sigma}\rangle}{\partial t}&=&\tilde{t}X_{\sigma}-U\langle c^{\dagger}_{i\bar{\sigma}}c_{i\bar{\sigma}}c^{\dagger}_{i\sigma}c_{j\sigma}\rangle+U\langle c^{\dagger}_{i\sigma}c_{j\sigma}c^{\dagger}_{j\bar{\sigma}}c_{j\bar{\sigma}}\rangle. \nonumber \\
& & \label{eq:fEOM}
\end{eqnarray}
Here $X_{\sigma}\equiv\langle c^{\dagger}_{i+1,\sigma}c_{j\sigma}\rangle+\langle c^{\dagger}_{i-1,\sigma}c_{j,\sigma}\rangle-\langle c^{\dagger}_{i\sigma}c_{j+1,\sigma}\rangle-\langle c^{\dagger}_{i\sigma}c_{j-1,\sigma}\rangle$.

When $U/\tilde{t}\le 1$, we implement the standard Hartree-Fock approximation by decomposing $\langle c^{\dagger}_{i\bar{\sigma}}c_{i\bar{\sigma}}c^{\dagger}_{i\sigma}c_{j\sigma}\rangle$ as $\langle c^{\dagger}_{i\bar{\sigma}}c_{i\bar{\sigma}}\rangle\langle c^{\dagger}_{i\sigma}c_{j\sigma}\rangle$.  This approximation closes the set of equations of motion and we can solve them with the initial condition $c_{ij}(t=0)=\sum_{p,p^\prime =1}^{N_p}(U_{int,L}^{\dagger})_{pi}(U_{int,L})_{jp^\prime}D^{L}_{int,pp^\prime}$ if $1\le i,j \le (N/2)$ and zero otherwise. Here $D^{L}_{int,pp^\prime}$ is the correlation matrix in the energy space of the left-half lattice of $H_{int}$ and $U_{int,L}$ is the corresponding unitary transformation. For general ratios of $N_{p\sigma}/N$, it is challenging to find $c_{ij}(t=0)$. However, in the case where initially there are two fermions (of opposite species) per site on the left-half lattice, Pauli exclusion principle requires $D^{L}_{int,pp^\prime}=\delta_{p,p^\prime}$ so $c_{ij}(t=0)=\delta_{ij}$ if $1\le i,j\le (N/2)$ and zero otherwise. We will focus on this case and investigate the effects of interactions. In our simulations we evolve the two species in a symmetric way so that $\langle c^{\dagger}_{i\sigma}c_{j\sigma}\rangle=\langle c^{\dagger}_{i\bar{\sigma}}c_{j\bar{\sigma}}\rangle$ during the evolution.

\begin{figure}
  \includegraphics[width=3.2in,clip]
{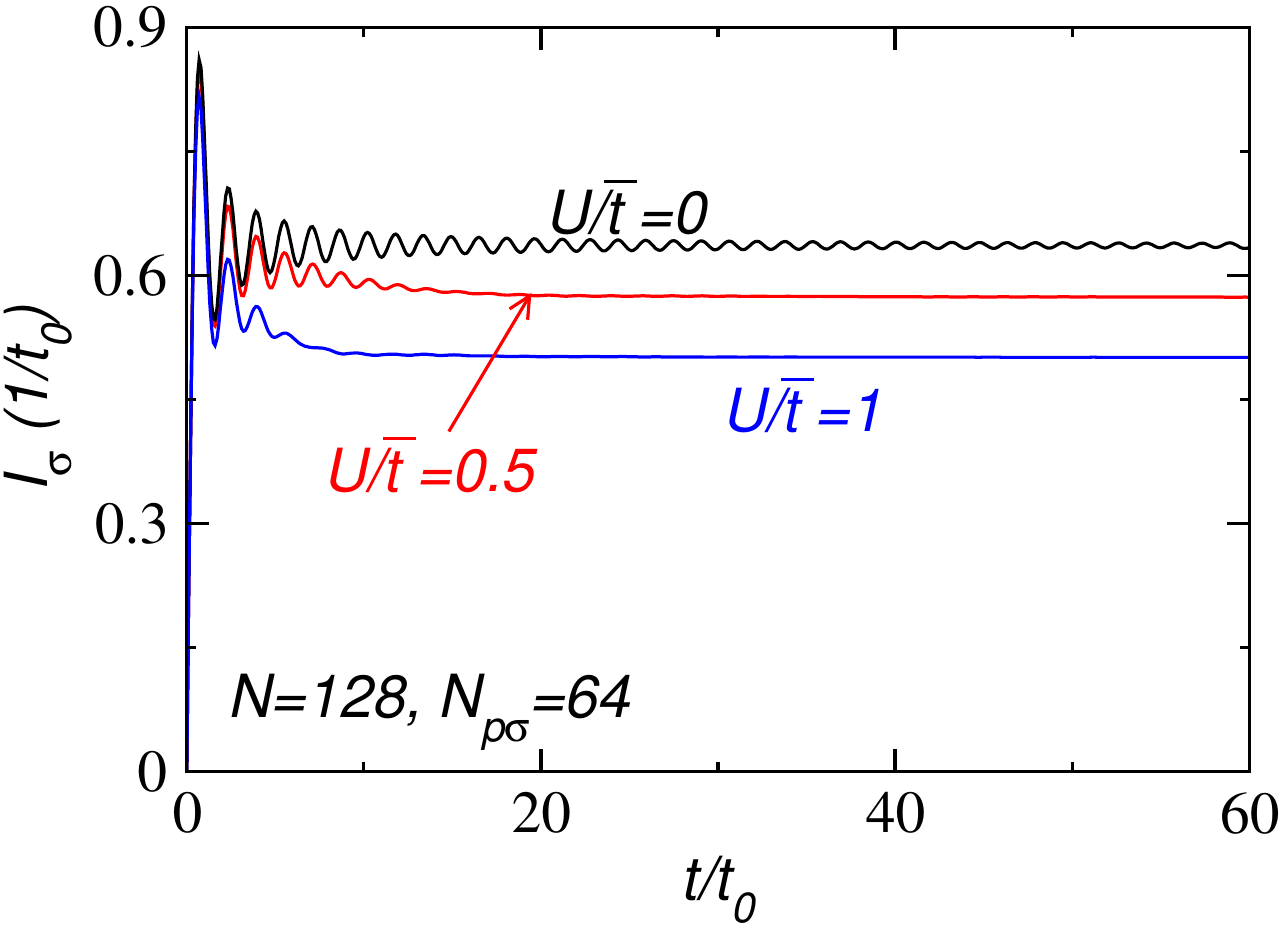}
  \caption{(Color online) Currents of one species of two-species fermions with interaction strength $U/\tilde{t}=0, 0.5, 1$ (labeled next to each curve). Here $N=128$ and $N_{p\sigma}=64$.}
\label{fig:I_int}
\end{figure}
Figure~\ref{fig:I_int} shows the current for $U/\tilde{t}=0, 0.5, 1$ with $N=128$ and $N_{p\sigma}=64$. One can see clearly that for each case there is a finite QSSC. Therefore the QSSC is robust and should be observable in either non-interacting single-species fermions or interacting two-species fermions if the interaction is moderate. The emergence of a QSSC in interacting fermions has also been found in different setups using the time-depenent density-matrix renormalization group method with fixed particle numbers \cite{LangerTDDMRG} and in interaction-induced transport, where time-dependent inhomogeneous interactions serve as an internal driving force for the current \cite{int_transport}.

There are two additional interesting effects shown in Fig.~\ref{fig:I_int}. The first one is that in the presence of interactions, the plateau of the QSSC is smoother and exhibits less noise. The interactions seem to rectify the QSSC and reduce the modulations on top of the plateau. The second is that the height of the plateau in the QSSC decreases as the interaction strength increases. This implies that even repulsive interactions can slow down the transport of fermions propagating from a localized region to an empty region. This is consistent with the experimental findings of Ref.~\cite{Blochtransport}, where fermions in a 3D optical lattice are initially localized due to a harmonic potential and then released from that potential. The experimental data suggest that interactions, whether repulsive or attractive, slow down the transport of fermions. Here we observe similar phenomena in 1D optical lattices.

In summary, we have demonstrated that the current of non-interacting ultra-cold fermions on 1D optical lattices exhibits a smooth crossover as one tunes the ratio of particle number and lattice size, $N_p/N$, from the empty-lattice limit to the infinite-volume limit and vice versa. The detailed study of this crossover is made possible in non-interacting systems using the MCF. While a finite current is not sustainable in the empty-lattice limit, a QSSC showing a plateau in $I(t)$ already emerges when the system size is reasonably large with finite filling. The threshold value of the filling where the QSSC emerges is the same for the two cases, and it corresponds to the deviation of Fermi energy from the bottom of the energy band. There is another threshold when the lattice size is too small and finite-size effects prohibits the formation of a QSSC. Moreover, we found that the QSSC also emerges in interacting two-component fermions with finite filling. Our study advances the understanding of the role of the thermodynamic limit in non-equilibrium isolated quantum systems, and we hope
it will motivate experiments in this direction.

We thank M. Zwolak for useful discussions. CCC acknowledges the support of the U. S. Department of Energy through the LANL/LDRD Program.
MD acknowledges support from the DOE grant DE-FG02-05ER46204 and UC Laboratories.

\bibliographystyle{apsrev}

\end{document}